\documentclass[10pt,aps]{revtex4}
\usepackage{amssymb}
\usepackage{latexsym}
\usepackage{epsfig}
\usepackage{float}
\usepackage{graphicx,epsfig, color}
\usepackage{graphicx}
\usepackage{subfigure}
\usepackage{hyperref}
\usepackage{amsmath}
\usepackage{xcolor}

\begin{document}
\title{\textbf{Stability of quark matter affected by the surface tension in a strong magnetic field}}
\author{Yu-Ying He and Xin-Jian Wen  }

\affiliation{Institute of Theoretical Physics, State Key Laboratory
of Quantum Optics and Quantum Optics Devices, Shanxi University,
Taiyuan, Shanxi 030006, China}

\begin{abstract}
The surface tension of quark matter in a strong magnetic field is investigated using a geometric approach. The interface between the hadronic phase and quark phase is determined by the Maxwell
construction of the first-order transition. When surface tension is included, the free energy per baryon is no longer a monotonic function of the chemical potential. Specifically, for smaller droplets, a larger chemical potential is required to achieve a stable phase. 
Moreover, we find that the surface tension does not increase monotonically with the magnetic field. Finally, it is shown that stable quark matter, both with and without surface tension, can exist at a specific magnetic field strength, which would provide favorable conditions for the experimental production of quark matter.

\end{abstract}



\maketitle

\section{Introduction}

Up to now, it is well known that strong magnetic fields are relevant in the early universe, magnetars, and non-central heavy ion collisions. They produce significant effects on the equation of state in compact stars and the QCD phase diagram. In particular, magnetic catalysis and inverse magnetic catalysis in the QCD chiral transition have been confirmed by lattice calculations \cite{Pisarski:1989cs, Schertler:1996tq}. The magnetic catalysis effect shows that a magnetic field enhances the spontaneous breakdown of chiral symmetry \cite{Klevansky:1989vi, Gusynin:1994xp,Shushpanov:1997sf, Ferrer:2000ed, Mueller:2015fka, Miransky:2002rp,
Inagaki:2003yi}. A more general result states that a constant magnetic field leads to the generation of a fermion dynamical mass \cite{Gusynin:1994re}. This surprising and unexpected inverse catalysis at finite temperatures was first predicted in lattice simulations \cite{Bali:2011qj, Preis:2010cq}. The corresponding theoretical interpretation is supported by investigations of the running coupling constant and/or the anomalous magnetic moment.
In all of these processes, the magnetic effect is demonstrated through changes in the order parameter of chiral symmetry. The inclusion of baryon density drastically changes the situation, and the transition becomes first-order. At high densities, the matter is always in the chirally symmetric phase. Furthermore, the formation of bubbles can be understood as a natural consequence of the first-order nature of the phase transition. The transition from the chiral symmetry broken hadron phase to the chiral symmetry restored quark phase is a first-order phase transition. The bubble formation of a quark phase in hadronic matter can be studied  through a nucleation process. These bubbles may consist of matter with restored chiral symmetry, while the surrounding hadronic matter remains in the chiral symmetry broken phase. Therefore, the study of bubbles composed of mixed phases is of great interest in the presence of a strong magnetic field.

Although the study of QCD phase transitions has made significant progress, the realistic occurrence of such transitions could be influenced by the finite volume of the system. Most model descriptions
of strong interactions at high density and low temperature suggest a
first-order phase transition for both the chiral and deconfinement transitions. Therefore, the interface composed of the two phases plays an important role. The surface tension \cite{Oertel:2008wr,Mintz:2012mz,Ke:2013wga,Gao:2016hks,Xia:2018wdj,Fraga:2018cvr,Lugones:2018qgu,Bogadi:2020sjy,Lugones:2020qll,Voskresensky:2022fzk,Mariani:2023kdu,Xia:2024juh} is defined as the excess free energy of a two-phase configuration containing a bubble wall, compared to a homogeneous, one-phase configuration, per unit wall surface area \cite{Endrodi:2021kur}. The investigation of surface tension and its impact on the stability of quark matter has been widely studied in the MIT bag model and the density-dependent quark mass model. However, their descriptions of interface tension lack detailed information about the phase transition. The quark quasiparticle model, as an extension of the bag model, has been developed to study the bulk properties of dense quark matter at finite density and temperature \cite{Bluhm:2008,Thaler:2003uz, Schertler:1996tq, He:2023gva}. This model provides a framework for understanding the medium effect on surface tension \cite{Wen:2010zz}. To describe the chiral phase transition and dynamical symmetry breaking, the Nambu-Jona-Lasinio (NJL) model is widely used in QCD-like investigations
\cite{Ferrer:2013noa,Avancini:2019wed,Allen:2015paa}. It effectively explains the surface tension at finite temperature, which is constructed by the mixed phase in the context of chiral symmetry breaking
\cite{Garcia:2013eaa}.

The surface tension is a characteristic property of coexisting phases. For example, the magnitude of the surface tension between a liquid drop and a gas phase plays a dominant role in the formation of raindrops. For QCD matter, the surface tension is a fundamental parameter for understanding finite-size effects and nucleation phase transitions
\cite{Huang:1990jf}. However, the surface tension between the true vacuum and the perturbative phase is often neglected in comparison to the confinement bag constant. For the mixed phase to occur, the surface tension at the interface separating the quark and hadron phases should be relatively small \cite{Maruyama:2007ey,Alford:2001zr}. A large value of the surface tension, such as 29.73 MeV$\cdot$fm$^{-2}$ , may be insufficient to stabilize quark matter during the phase transition. It has been proposed that strange matter could not have survived in the early Universe because it might boil and form bubbles of hadronic gas \cite{Alcock:1988br}. The study of surface tension should be considered a complementary approach to understanding QCD at finite density, as lattice QCD (LQCD) methods are insufficient to determine the structure of matter at larger chemical potentials. The
aim of this work is to investigate the surface tension of cold quark
matter in the NJL model under strong magnetic fields. We hope to explore the stability of quark matter modified by surface tension in a finite volume.

This paper is organized as follows. In Section \ref{sec:model}, we
present the thermodynamics of the magnetized quark matter in the NJL
model. The geometric approach is employed to define the surface
tension in Section \ref{sec:def}. The numerical results on the
stability of quark matter are presented in Section \ref{sec:result}.
The last section provides a brief summary.

\section{Thermodynamics of NJL model in a strong magnetic field}\label{sec:model}

The starting point of the NJL model as an effective
theory is the Lagrangian density, which models dynamical 
chiral symmetry breaking. The Lagrangian density of the two-flavor NJL model in a strong
magnetic field is given by \cite {Nambu:1961tp, Nambu:1961fr}:
\begin{equation}
{\mathcal{L}}_{NJL}=\bar{\psi}(i/\kern-0.5emD-m)\psi
+G[(\bar{\psi}\psi )^{2}+(\bar{\psi}i\gamma _{5}\vec{\tau}\psi
)^{2}]-\frac{1}{4}{F}_{\mu \nu }{F}^{\mu \nu },
\end{equation}%
where $\psi$ represents a flavor isodoublet (u and d quarks),
and $\vec{\tau}$ are isospin Pauli matrices. The covariant derivative
$D_{\mu }=\partial _{\mu
}-iQeA_{\mu }$ represents the coupling of the quarks to
the electromagnetic field, and a sum over flavor and color degrees of
freedom is implicit. Here, $Q$ =diag$(q_u,q_d)$=diag(2/3,-1/3) is the quark electric charge matrix in flavor space. The Abelian gauge field $A_{\mu }$ stands for the external magnetic
field $B$ aligned along the z-direction. In the mean-field approximation\cite {Ratti:2002sh, Buballa:1998pr}, the dynamical quark mass is directly dependent on the quark condensate\cite {Buballa:2003qv}:
\begin{equation}
M_i=m-2G\langle \bar{\psi}\psi \rangle,  \label{eq:gap}
\end{equation}
where the current masses $m_u=m_d=m$ are used, and the 
quark condensates include contributions from u and d quarks as
$\left\langle {\bar{\psi }\psi }\right\rangle=\displaystyle\textstyle\sum_{i=u,d}{\phi}_{i}$. 
For the SU(2) version, although the quark condensates for the flavors u and d in the
presence of a magnetic field are different due to their
different electric charges, the masses of the $u$ and $d$ constituent quarks are equal to each other in the isospin symmetric limit.
The constituent mass depends on both
condensates, and thus the same mass $M_u=M_d=M$ is
obtained for u and d quarks. The contribution from the
quark flavor i is given by \cite {Menezes:2008qt}:
\begin{eqnarray}
\phi _i=\phi _{i}^{\mathrm{vac}}+\phi _{i}^{\mathrm{mag}}+\phi _{i}^{\mathrm{med}}.
\end{eqnarray}

The terms $\phi _{i}^{\mathrm{vac}}$, $\phi _{i}^{\mathrm{mag}}$, and $\phi _{i}^{\mathrm{med}}$ represent the vacuum, magnetic field, and medium contributions to the quark condensation, respectively,\cite {Menezes:2009uc, Avancini:2011zz}
\begin{eqnarray}
\phi_i ^{\mathrm{vac}} &=&=-\frac{MN_{c}}{2\pi ^{2}}\left[ \Lambda \sqrt{%
	\Lambda ^{2}+M^{2}}-M^{2}\ln (\frac{\Lambda +\sqrt{\Lambda ^{2}+M^{2}}}{M})%
\right], \label{eq:convac}\nonumber
\hspace{-1.47cm}\\
\phi _{i}^{\mathrm{mag}} &=&-\frac{M|q_{i}|BN_{c}}{2\pi ^{2}}\left\{
\ln [\Gamma (x_{i})]-\frac{1}{2}\ln (2\pi)+x_{i}
-\frac{1}{2}(2x_{i}-1)\ln
(x_{i})\right\}, \\
\phi _{i}^{\mathrm{med}} &=&\sum_{k_{i}=0}a_{k_{i}}\frac{M|q_{i}|BN_{c}}{%
	2\pi ^{2}}\ln{\left ( \frac{\mu +\sqrt{{\mu }^{2}-{{M}_{n}^{2}}}}{{M}_{n}}\right )},
\end{eqnarray}
where $a_{k_{i}}=2-\delta_{k0}$ and $k_{i}$ are the degeneracy label
and the Landau quantum number, respectively. The dimensionless
quantity $x_i$ is defined as $x_i=M^{2}/(2|q_{i}|B)$. ${M}_{n}=\sqrt{{M}^{2}+2{k}_{i}\left | {q}_{i}\right |B}$.

The total thermodynamic potential density in the mean-field
approximation is:
\begin{equation}\label{omega}\Omega=\frac{(M-m_0)^2}{4G}+\sum_{i=u,d}\Omega_i,
\end{equation}
where the first term is the interaction term. In the framework of the magnetic-field-independent regularization
\cite {Ebert:1999ht, Vdovichenko:2000sa, Frolov:2010wn}, the thermodynamic potential $\Omega_i$ is usually divided into three terms:
\begin{eqnarray}
\Omega_i=\Omega_i^\mathrm{vac}+\Omega_i^\mathrm{mag}+\Omega_i^\mathrm{med}.
\end{eqnarray}

The vacuum contribution to the thermodynamic potential is
\begin{eqnarray}
\Omega_i ^{\mathrm{vac}}=\frac{N_{c}}{8\pi ^{2}}\left[ M^{4}\ln
(\frac{\Lambda +\epsilon _{\Lambda }}{M})-\epsilon _{\Lambda
}\Lambda (\Lambda ^{2}+\epsilon _{\Lambda }^{2})\right],
\end{eqnarray}
where $\epsilon _\Lambda=\sqrt{\Lambda^2+M^2}$. The ultraviolet divergence in the vacuum part $\Omega_i^{\mathrm{vac}}$ of the thermodynamic potential is removed by the momentum cutoff. The magnetic field and medium contributions are respectively\cite {Menezes:2015fla, Menezes:2009uc,
	Avancini:2011zz}:
\begin{eqnarray}
\Omega _{i}^{\mathrm{mag}}&=&-\frac{N_{c}(|q_{i}|B)^{2}}{2\pi
^{2}}\left[
\zeta^{\prime}(-1,x_i)-\frac{1}{2}(x_{i}^{2}-x_{i})\ln (x_{i})+\frac{%
x_{i}^{2}}{4}\right],\\
 \Omega _{i}^{\mathrm{med}}
&=&-\sum_{k_{i}=0}a_{k_{i}}\frac{|q_{i}|BN_{c}}{4\pi ^{2}}\left \{ \mu \sqrt{{\mu }^{2}-{M}_{n}^{2}}-{M}_{n}^{2}\ln\left (\frac{\mu +\sqrt{{\mu }^{2}-{M}_{n}^{2}}}{{M}_{n}} \right )\right \},
\end{eqnarray}
where $\zeta (a,x)=\sum_{n=0}^{\infty }\frac{1}{(a+n)^{x}}$ is the
Hurwitz zeta function \cite{Menezes:2008qt}.

\section{THE surface tension in a geometric approach}\label{sec:def}
In literature, surface tension has been investigated using two main approaches. 
One is the multiple reflection expansion approximation
\cite{Shao:2006gz}, and the other is the geometric approach
\cite{Pinto:2012aq,Garcia:2013eaa}. The latter method incorporates 
more information about the interaction. In this section, we review
the geometric approach for calculating surface tension, which was
first proposed in Ref. \cite {Cahn:1959tez,Ravenhall:1983bdb}. For any subcritical
temperature $T<{T}_{c}$, two phases with different densities can
reach thermodynamic equilibrium in phase coexistence. When these two
phases come into contact, a mechanically stable interface forms
between them, and the surface tension ${\gamma }_{T}$ characterizes this interface. 
As the temperature gradually increases from 0 to ${T}_{c}$, the two densities coincide at
the critical temperature ${T}_{c}$, and the surface tension eventually vanishes.

The free energy density ${f}_{T}\left(\rho\right)$ can be expressed
as: ${f}_{T}\left(\rho
\right)={\mathcal{E}}_{T}\left(\rho\right)-T{s}_{T}\left(\rho\right
)={\mu}_{T}\left(\rho\right )\rho-{P}_{T}\left(\rho\right)$, which
includes the fundamental relation of the energy density
$\mathcal{E}_{T}$ and the pressure ${P}_{T}$. When ${f}_{T}\left(\rho\right)$ exhibits local concavity, it
indicates that the system in that region is unstable. As a result, the system
will tend to decompose into two phases: a low-density phase ${\rho}_{1}\left(T\right)$
and a high-density phase ${\rho}_{2}\left(T\right)$. To achieve thermodynamic 
equilibrium between these two phases, a common tangent of the free
energy at densities ${\rho}_{1}$ and ${\rho}_{2}$ corresponds to a
common chemical potential and defines the Maxwell free energy
density ${f}_{T}^{M}\left(\rho\right)$. Consequently, the cost of the 
free energy density, defined as
$\Delta{f}_{T}\left(\rho\right)\equiv{f}_{T}\left
(\rho\right)-{f}_{T}^{M}\left(\rho\right)\geqslant 0$, is non-negative.

The interface tension ${\gamma }_{T}$ is given by 
\begin{eqnarray}
{\gamma}_{T}=a\int_{{\rho}_{1}}^{{\rho}_{2}}\frac{d\rho}{{\rho}_{g}}{\left [2{\mathcal{E}}_{g}\Delta{f}_{T}\left(\rho \right)\right]}^{\frac{1}{2}},
\end{eqnarray}
where ${\rho}_{g}$ is a characteristic value of the density, ${\mathcal{E}}_{g}$ is a characteristic value of the energy density, and the parameter $a\approx 1/{m}_{\sigma }\approx 0.33$ fm.

In the zero-temperature limit, the interface tension is controlled by the magnetic field. The free energy density is given by the following equation:
\begin{eqnarray}
f=\mu\rho-P=\mu\rho+\Omega,
\end{eqnarray}
\begin{eqnarray}
\Delta f=\left (\Omega-{\Omega}_{M}\right)+\rho\left (\mu-{\mu}_{M}\right),
\end{eqnarray}
where $P$ is the pressure, $\Omega$ is the thermodynamic potential density, $\mu$ is the quark chemical potential, and $\rho$ is the quark number density. Here, ${\Omega }_{M}$ and ${\mu }_{M}$ are the thermodynamic potential density and quark chemical potential of the Maxwell structure, respectively. A more detailed discussion can be found in Ref. \cite {Garcia:2013eaa,Randrup:2009gp}

For the case of a finite spherical droplet, we assume spherical
symmetry with radius $R$. The total free energy is expressed as a sum
of the volume and surface terms:
\begin{eqnarray}
f=\left(\mu\rho+\Omega\right)V+\gamma S,
\end{eqnarray}
where $V$ is the volume and $S$ is the surface area. Correspondingly, the free energy per baryon, including the coexisting phases, is given by
\begin{eqnarray}
\frac{f}{A}=\frac{\Omega}{{\rho}_{B}}+3\mu+\frac{3\gamma }{{\rho}_{B} R},
\end{eqnarray}
where $A$ is the baryon number, and the baryonic density ${\rho}_{B}$
can be written as\cite {Menezes:2008qt}
\begin{eqnarray}
{\rho}_{B}=\frac{1}{3}\left({\rho }_{u}+{\rho}_{d}\right)=\displaystyle\sum_{i=u}^{d}\displaystyle\sum_{{k}_{i}=0}^{{k}_{i}^{max}}{a}_{{k}_{i}}\frac{\left|{q}_{i}\right|B{N}_{c}}{6{\pi}^{2}}\sqrt{{\mu}^{2}-2{k}_{i}\left | {q}_{i}\right |B-{M}^{2}}.
\end{eqnarray}

\section{Numerical result and conclusion}\label{sec:result}

In this section, we present numerical results for the surface
tension between the chiral broken phase and the chiral restored
phase. Furthermore, we demonstrate that the stability of 
finite-size quark matter is influenced by the surface tension. In
our model, the following parameters are adopted for the evaluation:
${m}_{u}={m}_{d}=5.6$ MeV, $\Lambda=631$ MeV, $G = 2.19/{\Lambda
}^{2}$.

\begin{figure}[H]
    \centering
    \begin{minipage}{8cm}
        \includegraphics[width=8cm]{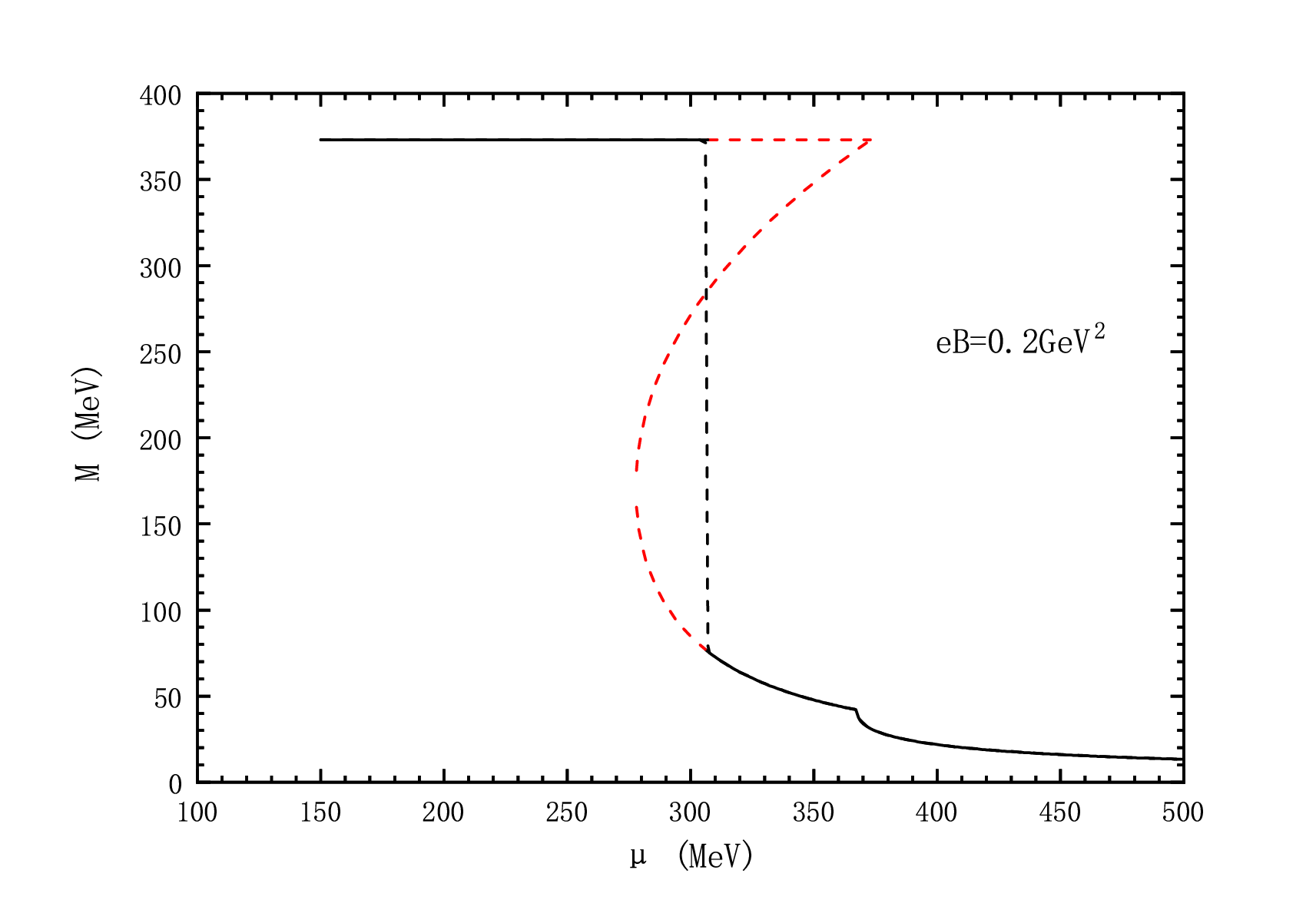}
    \end{minipage}
\caption{\label{fig:wide} The variation of the constituent quark
mass $M$ with the quark chemical potential $\mu$ is shown. The red dashed line and the black solid line represent the unstable and stable solutions, respectively. The black dashed line represents the
Maxwell construction. }\label{fig1}
\end{figure}

By solving the gap equation (\ref{eq:gap}), the obtained effective mass is used to
determine the thermodynamic potential density. The free energy
density $f$ can then be calculated to determine the difference $\Delta
f$, which is a crucial ingredient in the evaluation of surface tension. 
For more detailed references, see Refs. \cite{Garcia:2013eaa,Randrup:2009gp}. 
In Fig. \ref{fig1}, the variation of constituent quark mass $M$ is 
shown as a function of quark chemical potential $\mu$ at $T\to$ 0 for $eB=0.2$ GeV$^2$. 
The black solid line represents the stable solutions of the dynamical mass $M$, 
while the red dashed line represents the unstable solutions in 
the first-order transition. The Maxwell line is determined by 
the black dashed line. The dynamical quark mass remains constant in the vacuum
and becomes multivalued as the chemical potential increases up
to $\mu\backsim$ 280 MeV, which marks the beginning of a 
first-order transition. The purely chiral symmetry restored 
phase emerges when the chemical potential exceeds $\mu=374$ MeV. 
Thus, the quark chemical potential range of 280 MeV to 374 MeV 
corresponds to the coexisting phases at the magnetic field $eB=0.2$ GeV$^2$.
There exists a kink for the dynamical quark mass at
around $\mu=370$ MeV. This is produced by the Landau level
quantization effect. The strong magnetic field causes the
discretization of quark energy levels. When the Fermi surface
crosses a Landau level, it leads to a sudden change in the density
of states and the effective mass.

\begin{figure}[H]
    \centering
    \begin{minipage}{8cm}
        \includegraphics[width=8cm]{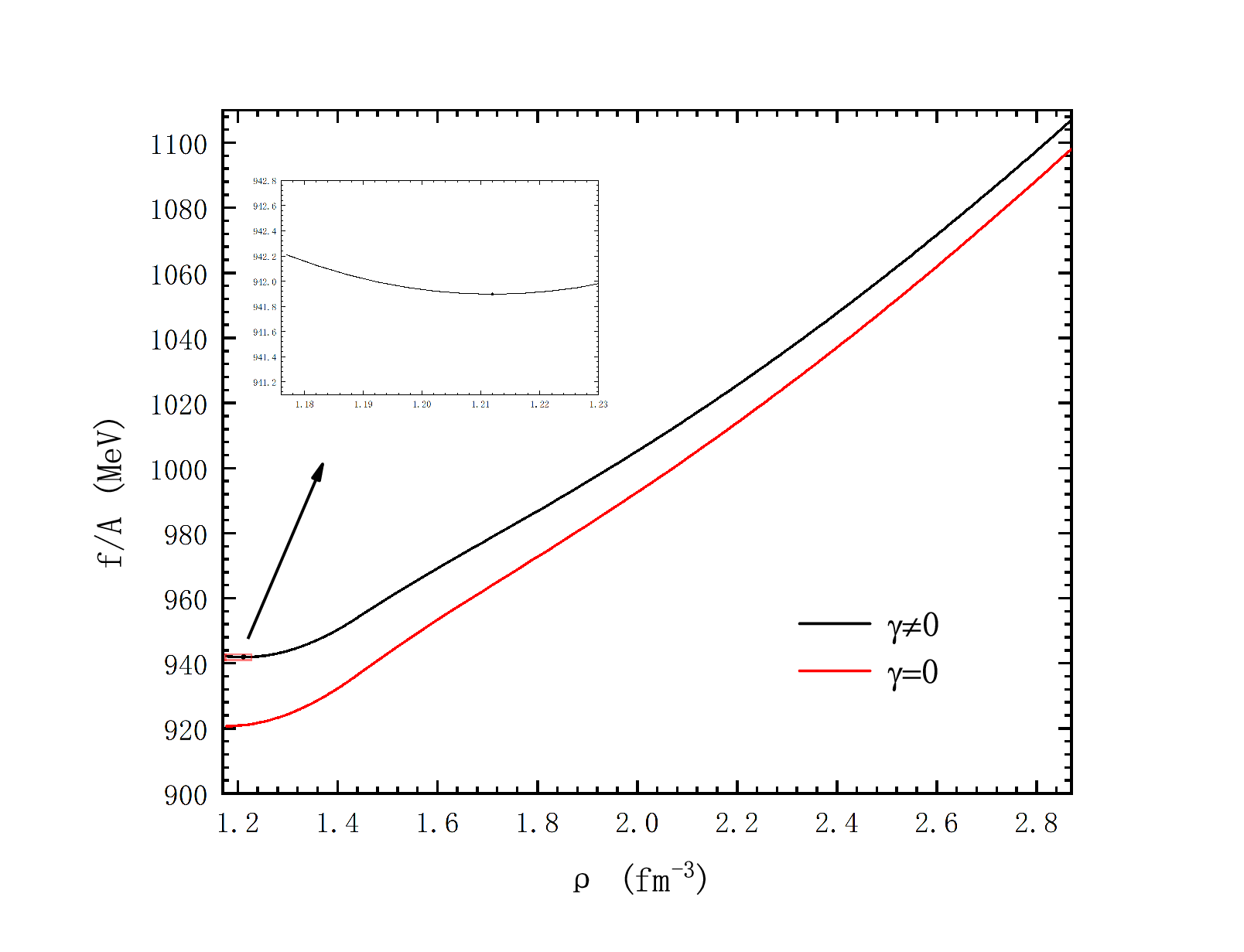}
    \end{minipage}
    \caption{\label{fig:wide}The free energy per baryon $f/A$ as a function of the quark number density $\rho$ is shown, both with and without surface tension, at $eB=0.2$ GeV$^2$. }\label{fig2}
\end{figure}

\begin{figure}[H]
    \centering
    \begin{minipage}{8cm}
        \includegraphics[width=8cm]{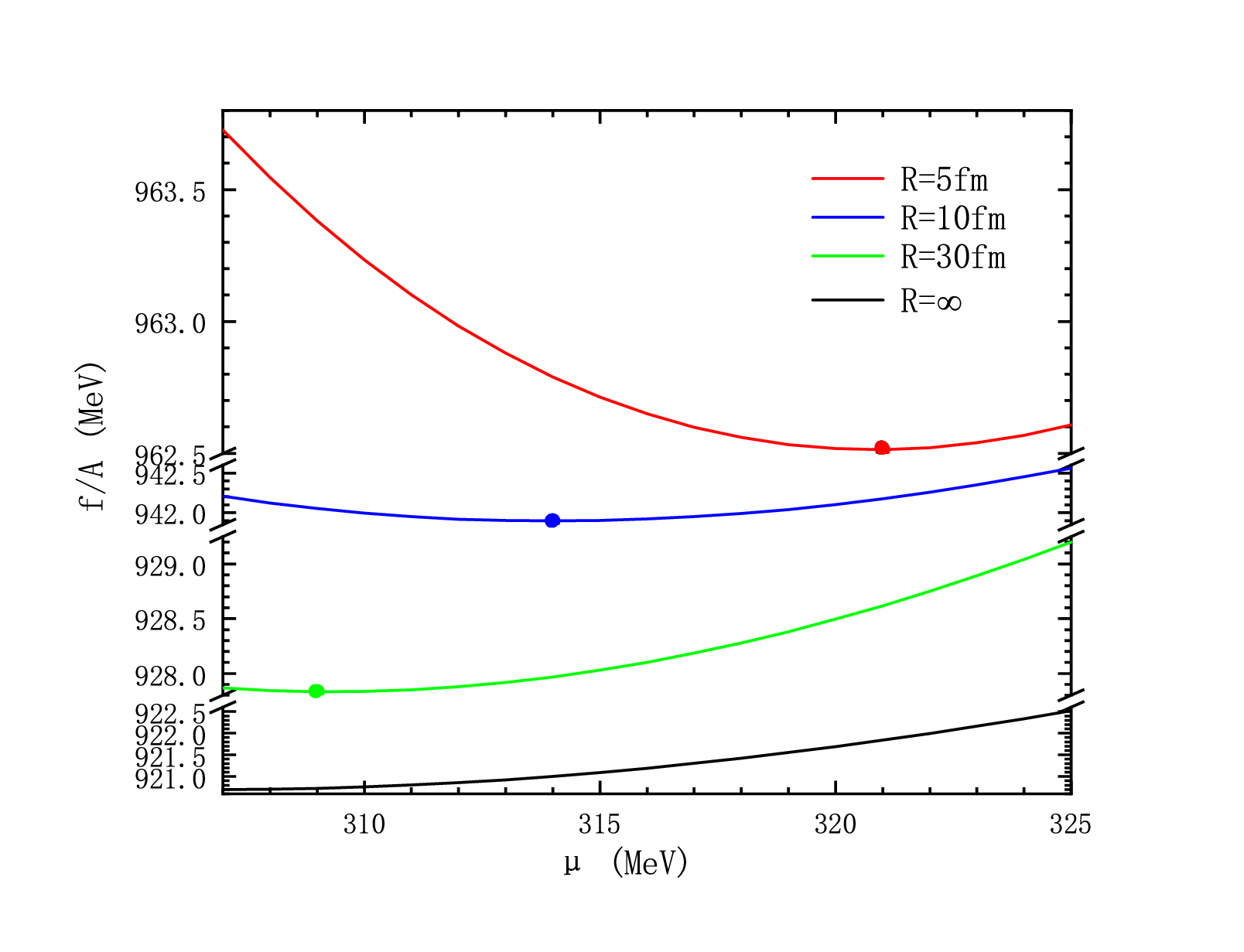}
    \end{minipage}
\caption{\label{fig:wide}The free energy per baryon $f/A$ as a
function of the quark chemical potential $\mu$ is shown for $\gamma\neq
0$ (at $eB=0.2$ GeV$^2$) and radii $R$=5, 10, and 30 fm. The monotonic line at the bottom corresponds to the bulk matter case ($R=\infty$). }\label{fig3}
\end{figure}

\begin{figure}[H]
    \centering
    \begin{minipage}{8cm}
        \includegraphics[width=8cm]{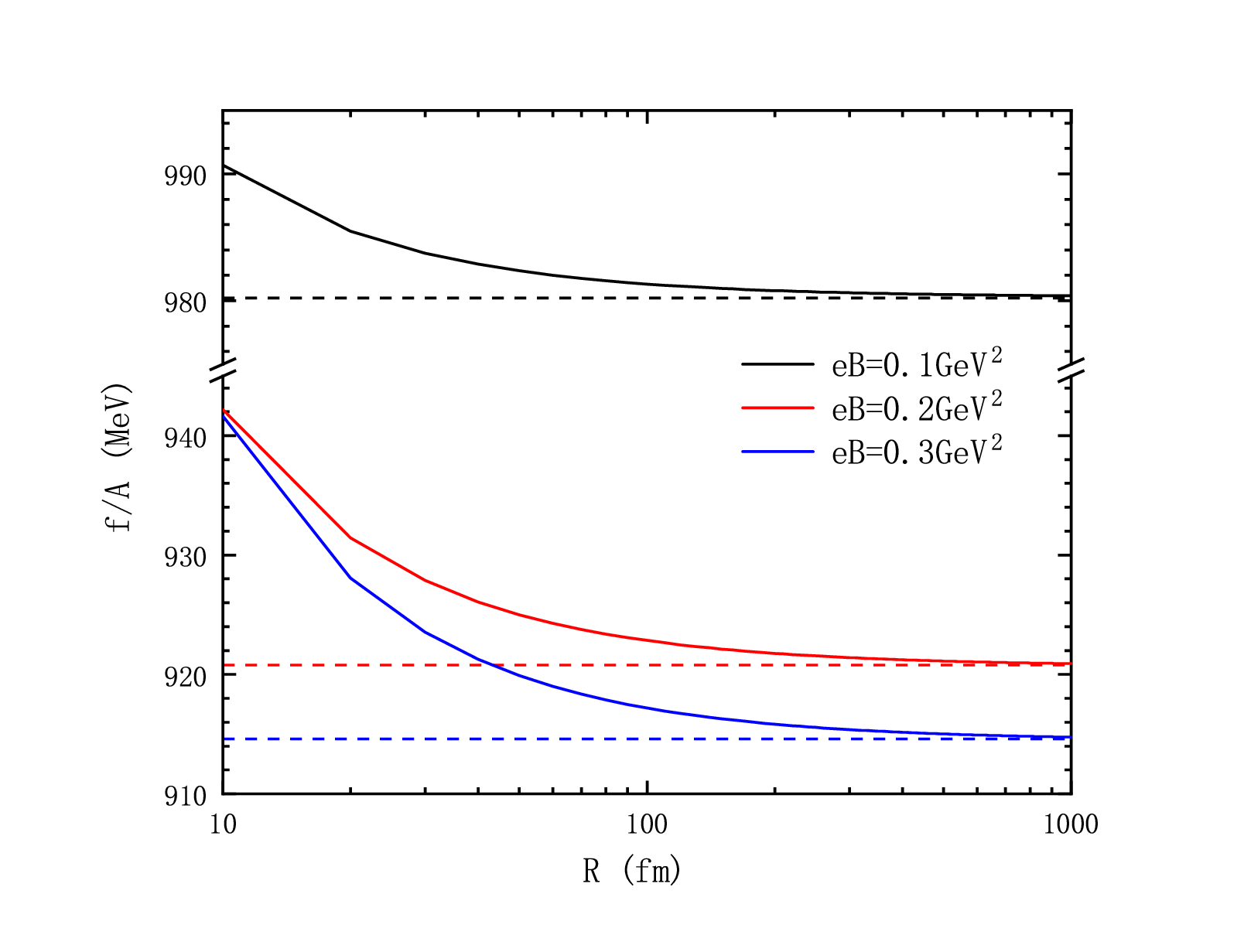}
    \end{minipage}
    \caption{\label{fig:wide}The free energy per baryon as a function of the radius of a finite droplet is shown for $eB=$ 0.1, 0.2, and 0.3 GeV$^2$, with the radius $R$ varying from 10 fm to 1000 fm. }\label{fig4}
\end{figure}

In our study, a spherical system composed of the chiral broken and
restored phases is assumed to investigate the finite-size effect.
The electric field configurations determine the droplet size and the
geometry of structures \cite{Voskresensky:2002hu}. When discussing 
relatively short-distance effects, one may disregard spatial changes 
in sphericity due to the presence of a strong magnetic field. The 
free energy of the system can be expressed as the sum of the bulk 
matter term and the surface term. The free energy per baryon $f/A$ is shown as a
function of the density $\rho$ for a given radius $R=10$ fm in Fig.
\ref{fig2}. It is evident that, when surface tension is considered, 
the free energy per baryon initially decreases and then increases 
as the density increases, as shown in the inset of Fig 2, where 
the solid dot indicates the position of the minimum. Without surface 
tension, the free energy per baryon, marked by the red line, remains 
a monotonically increasing function of density. Moreover, the inclusion 
of surface tension always results in a higher free energy per baryon 
compared to that of bulk matter. Consequently, it is concluded that 
surface tension reduces the thermodynamic stability of quark matter.
Spherical radii of $R=5, 10, 30$ fm are adopted for comparison.
In Fig. \ref{fig3}, the free energy per baryon $f/A$ is shown as a
function of quark chemical potential. When the surface tension is
taken into account, the non-monotonic behavior becomes clearly 
evident and is more pronounced than that in Fig. \ref{fig2}. 
The free energy per baryon initially decreases and then increases 
as the chemical potential increases. The minimum values of $f/A$, 
marked by solid dots in red, blue, and green, are reached at the 
following points: $R=5$ fm and $\mu= 321$ MeV (red dot), $R=10$ fm 
and $\mu=314$ MeV (blue dot), and $R=30$ fm and $\mu=309$ MeV (green dot). 
As the radius increases, the position of the minimum free energy 
per baryon shifts toward lower chemical potentials. For sufficiently 
large radii, the behavior approaches the monotonic curve of bulk matter, 
indicated by the black line at the bottom. For smaller radii, a larger 
chemical potential is required to stabilize the spherical volume. 
In general, it is confirmed that surface tension significantly increases 
the free energy per baryon due to the interface contribution.
Fig. \ref{fig4} illustrates the free energy per baryon as a function
of the finite droplet radius $R$ at fixed magnetic fields of 
$eB=$0.1, 0.2, and 0.3 GeV$^2$. The horizontal dashed lines represent 
the corresponding values for bulk quark matter. As the radius increases, the free
energy per baryon gradually decreases and eventually approaches the
constant value of bulk matter. It is evident that surface tension 
has a more significant effect on the increase of the free energy per 
baryon at small droplet sizes. Consequently, the stability of the 
quark matter system is reduced by the surface energy. By comparing 
the results for different magnetic fields, it is found that the effect 
of surface tension is more pronounced for stronger magnetic fields. For
the magnetic field $eB=0.3$ GeV$^2$, the free energy per baryon increases 
by approximately 3 percent compared to that of bulk matter.
To explore the effects of the magnetic field and surface tension on the stability of quark matter, we investigate the variation of the free energy with respect to the magnetic field. Fig. 5 presents the free energy per baryon as a function of the magnetic field on the left axis. The free energy with and without surface tension is marked by the black and the red lines, respectively. These two lines both reach their minimum values at a magnetic field strength of 0.26 GeV$^2$, after which they begin to rise steadily, indicating the existence of a stable phase at an appropriate magnetic field strength. 
The difference between these two lines increases with the magnetic field strength in the range of 0.04 {GeV}$^{2}$ to 0.26 {GeV}$^{2}$. When the magnetic field strength exceeds 0.26 GeV$^2$, the difference between the two lines gradually approaches a constant. In the magnetic field range of 0.04 {GeV}$^{2}$ to 0.06 {GeV}$^{2}$, the near overlap of the two lines is due to the very small surface tension at these magnetic fields, as indicated by the black arrow in the figure.
 
The surface tension is shown on the right axis in Fig. 5. It exhibits a non-monotonic dependence on the magnetic field, reaching its minimum value at 0.06 GeV$^2$. 
The blue dots illustrate the variation of surface tension with magnetic field as reported in Ref. [\cite{Garcia:2013eaa}]. Compared to our results, which are represented by the black dotted line, the small discrepancy is led to by the parameter set in the model. Furthermore, the trends observed in the two datasets demonstrate a notable alignment to some extent.
\begin{figure}[H]
    \centering
    \begin{minipage}{8cm}
        \includegraphics[width=8cm]{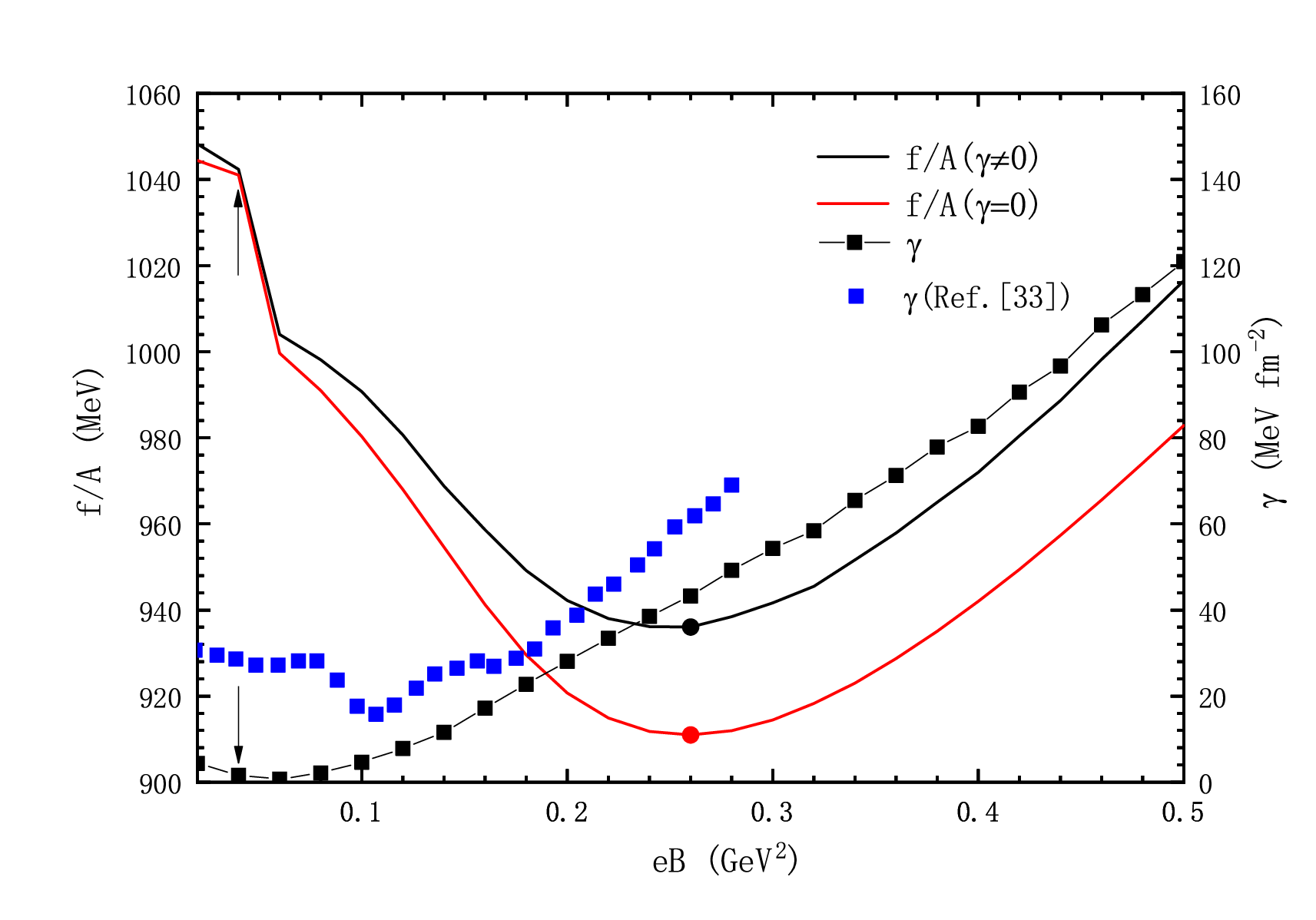}
    \end{minipage}
    \caption{\label{fig:wide}The free energy per baryon on the left axis and the surface tension on the right axis are shown as functions of the
    	magnetic field. The blue points correspond to the variation of surface tension with the magnetic field, as reported in Ref. [\cite{Garcia:2013eaa}].  }\label{fig5}
\end{figure}
\section{Summary}
In this work, we have investigated the surface tension related to
the first-order chiral phase transition of two-flavor magnetized
quark matter within the NJL model. The surface tension is defined
through a geometric approach. The Maxwell construction is required 
to determine the equation of state in the coexisting phase. By assuming a
spherical volume for the system, we have analyzed the variation of
the free energy per baryon with respect to quark number density and quark
chemical potential. It has been shown that the free energy per
baryon depends on the finite droplet radius and magnetic field
strength. Our results indicate that the free energy per baryon, 
when including surface tension, is not a monotonic function of 
increasing baryon density and quark chemical potential. The stable 
chemical potential is determined by the minimum free energy per 
baryon, which depends on the droplet radius. Specifically, smaller 
droplets in the coexisting phase require a larger chemical potential. 
Moreover, the free energy per baryon with surface tension is always 
higher than that of bulk matter. This suggests that the inclusion 
of surface tension inevitably reduces the stability of quark matter. 
It is also shown that the free energy per baryon decreases as the 
spherical radius increases, eventually approaching the bulk limit.

The surface tension coefficient is a significantly increasing 
function of the magnetic field for 0.06 GeV$^2$$<$eB$<$0.26 GeV$^2$. However, at $eB=0.06$ GeV$^2$, the surface tension is minimal, and consequently, 
the finite size has a weak influence on the free energy. Finally, it has been demonstrated that 
stable quark matter exists at an appropriate magnetic field, 
regardless of whether surface tension is included. We argue 
that such a proper magnetic field condition is favorable for 
producing the quark-gluon phase in experiments.

 \acknowledgments{ The
authors would like to thank support from the National Natural
Science Foundation of China under the Grant Nos. 11875181 and 12047571. This work was also sponsored by the Fund for Shanxi
"1331 Project" Key Subjects Construction.}

\end{document}